\RequirePackage[2020-02-02]{latexrelease}
\documentclass[singlecolumn,preprintnumbers,amsmath,amssymb]{revtex4}
\usepackage{graphicx}
\usepackage{dcolumn}
\usepackage{bm}
\usepackage{subfigure}

\begin{document}

\preprint{APS/123-QED}

\title{New Measurement of the Hoyle State Radiative Transition Width}

\author{T. K. Rana$^{1,2}$\footnote{Email:tapan@vecc.gov.in}, Deepak Pandit$^{1,2}$\footnote{Email:deepak.pandit@vecc.gov.in}, S. Manna$^{1,2}$, S. Kundu$^{1,2}$, K. Banerjee$^{1,2}$, A. Sen$^{1,2}$, R. Pandey$^{1}$,\\ G. Mukherjee$^{1,2}$, T. K. Ghosh$^{1,2}$, S. S. Nayak$^{1,2}$, R. Shil$^{3}$, P. Karmakar$^{1,2}$, K. Atreya$^{1,2}$, K. Rani$^{1}$, \\ D. Paul$^{1,2}$, Rajkumar Santra$^{1}$, A. Sultana$^{1,2}$, S. Basu$^{1,2}$, S. Pal$^{1,2}$, S. Sadhukhan$^{1,2}$, Debasish Mondal$^{1}$, \\ S. Mukhopadhyay$^{1,2}$, Srijit Bhattacharya$^{4}$, Surajit Pal$^{1}$, Pankaj Pant$^{1}$, Pratap Roy$^{1,2}$, Sk M. Ali$^{1}$,\\ S. Mondal$^{5}$, A. De$^{6}$, Balaram Dey$^{5}$, R. Datta$^{7}$, S. Bhattacharya$^{1,\ddag}$, C. Bhattacharya$^{1}$\footnote{superannuated.}}

\affiliation{%
\mbox{$^{1}$Variable Energy Cyclotron Centre, 1/AF, Bidhan Nagar, Kolkata-700064, India}\\
\mbox{$^{2}$Homi Bhabha National Institute, Mumbai – 400094, India} \\
\mbox{$^{3}$Siksha Bhavana, Visva-Bharati, Santiniketan-731235, India} \\
\mbox{$^{4}$Deptartment of Physics, Barasat  Government College, Kolkata – 70124, India} \\
\mbox{$^{5}$Department of Physics, Bankura University, Bankura - 722155, India} \\
\mbox{$^{6}$Department of Physics, Raniganj Girls’ College, Raniganj 713358, India}\\
\mbox{$^{7}$Nuclear Physics Division, Bhabha Atomic Research Centre, Mumbai - 400085, India} \\
}%

\date{\today}

\begin{abstract}
 The radiative decay of the Hoyle state serves as the gateway to the production of heavier elements in a stellar environment. Here, we present an  exclusive measurement of electric quadruple (E2) transitions of the Hoyle state to the ground state of $^{12}$C through the $^{12}$C(p, p$^\prime$$\gamma$$\gamma$)$^{12}$C reaction. A triple coincidence measurement yields the radiative branching ratio $\Gamma_{rad}$/$\Gamma$ = 4.01 (30) $\times$ 10$^{-4}$. This result was corroborated by an independent experiment based on the complete kinematical measurement $via$ $^{12}$C(p, p$^\prime$)$^{12}$C reaction, yielding a consistent result of $\Gamma_{rad}$/$\Gamma$ = 4.04 (30) $\times$ 10$^{-4}$. Combining our results with the currently adopted values of $\Gamma_{\pi}(E0)/\Gamma$ and $\Gamma_{\pi}(E0)$, the radiative width of the Hoyle state is determined to be
  3.75 (40) $\times$ 10$^{-3}$ eV. It is important to note that our finding do not align with a recently reported 34$\%$ increase in the radiative decay width of the Hoyle state but is consistent with the currently accepted value.
\end{abstract}

\maketitle


 The Hoyle state, the second excited state of $^{12}$C at an excitation energy of 7.65 MeV, sits just 380 keV above the 3$\alpha$ decay threshold (7.27 MeV), plays the most crucial role in the stellar nucleosynthesis of carbon and all heavier elements above $^{4}$He  ~\cite{hoyle,dubnar}. The presence of a resonance (due to the Hoyle state)  has catapulted the production of carbon through sequential 3$\alpha$ ($\alpha$+$\alpha$ $\rightleftharpoons$ $^{8}$Be; $^{8}$Be+$\alpha$ $\leftrightharpoons$ $^{12}$C) process to $\sim$ 7-8 order in magnitude compared to the normal non-resonant 3$\alpha$ process~\cite{hoyle}. This enhancement has helped to provide a satisfactory quantitative explanation of the solar abundance of carbon as well as heavier elements produced through the next stages of the fusion process ~\cite{BFH, fowler}. Even small changes in the rates of 3$\alpha$ $\leftrightharpoons$ $^{12}$C reaction can strongly influence the elemental abundance as well as the subsequent stages of stellar evolution ~\cite{beard,shilun,wanajo}. For instance, the rate of carbon formation $via$ the 3$\alpha$ reaction competes with the rate of carbon depletion in the next stage of the fusion chain, namely $^{12}$C($\alpha$, $\gamma$)$^{16}$O, to maintain the C/O ratio ($\sim$ 0.6 in the solar system), which fundamentally shapes the lifecycle of a star \cite{BFH,woosley}. Hence, the formation of the Hoyle state and its decay, as well as the destruction of carbon in $^{12}$C($\alpha$, $\gamma$)$^{16}$O reaction should be critically studied with utmost precision. In recent times, several high-precision studies have been made on the effect of the structure of the Hoyle state on the 3$\alpha$ decay rate ~\cite{rana1, Dell, Smith} to arrive at a general consensus; however, some inconsistencies still prevail regarding the decay of Hoyle state leading to carbon ground state (forming stable carbon), which necessitates further attention ~\cite{freer}.

 The stable $^{12}$C is now well known to be produced through the Hoyle state $via$ the triple-$\alpha$ reaction. Notably, the predominant decay pathway of the Hoyle state, denoted as $\Gamma_{\alpha}$, leads back to the same entrance channel (3$\alpha$) $via$ sequential decay, i.e.,$^{12}$C $\leftrightharpoons$ $^{8}$Be+$\alpha$; $^{8}$Be$ \rightleftharpoons$ $\alpha$+$\alpha$ \cite{freer}. Additionally, there may be a weak direct decay component of 3$\alpha$ ($^{12}$C $\rightarrow$ $\alpha$ + $\alpha$ + $\alpha$), which, although not yet experimentally observed (has a measured upper limit at 0.019$\%$), holds significant importance in elucidating the distinctive structure of the Hoyle state \cite{rana1,mfd}. Nevertheless, there is an extremely small branch ($\sim$ 0.04 $\%$) to form stable $^{12}$C by radiative transition ($\Gamma_{rad}$) to the ground state. The $\Gamma_{rad}$ is the sum of the cascade-$\gamma$ decay, $\Gamma_\gamma$(E2), through first 2$^+$ state of $^{12}$C ($>$ 98$\%$), internal pair formation, $\Gamma_\pi$(E0), directly to the ground state ($\sim$ 1.9$\%$) and the pair production, $\Gamma_\pi$(E2), ($<$ 0.1$\%$). Besides radiative decay, neutron-upscattering could potentially contribute to the formation of stable $^{12}$C; however, the observed neutron-upscattering enhancement is notably smaller than expected, indicating its significance might be confined only to certain astrophysical contexts like neutron star mergers \cite{nupscat}. Hence, the prevailing mechanism through which stable $^{12}$C forms is mainly $via$ the cascade-$\gamma$ decay of the Hoyle state. In astrophysical terms, the rate of the triple-$\alpha$ reaction can be expressed as~\cite{rolf}, R = C T$^{-\frac{3}{2}}(\Gamma_{\alpha} \Gamma_{rad}/\Gamma) exp(-Q/kT)$,
where, C is the proportionality constant, $ Q $ is the energy released in three $\alpha$ decay, $ T $ is the
temperature of the stellar environment and  $\Gamma$ is the total decay width of the Hoyle state ($\Gamma$  = $\Gamma_{\alpha}$ + $\Gamma_{rad}$). As $ \Gamma_{rad} \ll \Gamma_{\alpha} $, $ \Gamma \simeq \Gamma_{\alpha} $ and the 3$\alpha$ reaction rate is completely determined by the value of  $\Gamma_{rad}$. Experimentally, $\Gamma_{rad}$ can only be determined through the product of three independently measured quantities expressed as

\begin{equation}\label{2}
\Gamma_{rad}=\left [\frac{\Gamma_{rad}}{\Gamma}\right] \left [\frac{\Gamma}{\Gamma_\pi(E0)}\right] \left[\Gamma_\pi(E0)\right]
\end{equation}

Numerous experiments have been conducted to determine the aforementioned quantities ~\cite{freg, cran, gud, cran1, streh, streh1,cran2, chernykh, selove, alburger, obst, alburger1, robertson, eriksen, alburger2, seeger, obst1, hall, chamber, david, mak,markh} and the recommended values are compiled in tabulated form in Ref~\cite{kelley}. The current adopted values of $\Gamma_{rad}/\Gamma$, $\Gamma_\pi$(E0)/$\Gamma$  and $\Gamma$$_\pi$(E0) are 4.16(11) $\times$ 10$^{-4}$,  6.7 (6) $\times$ 10$^{-6}$ and 62.3 (20) $\mu$eV, respectively ~\cite{kelley}. These values have been utilized thus far for estimating the synthesis of all elements $via$ triple-$\alpha$ reaction rate. More recently, a new measurement of $\Gamma_{rad}$ has been performed, reporting a 34 $\%$ higher than the previous measurements~\cite{kibedi}. This large increase in the radiative decay width implies a significant change in the production cross-section of $^{12}$C, which has profound impact on the whole nucleosynthesis process as well as the theoretical model calculations ~\cite{woosley}. Consequently, further verification or confirmation through experiments with higher statistical significance and/or employing new measurement techniques is warranted. In this Letter, two independent experiments have been conducted to measure the radiative decay width of the Hoyle state: one utilizing triple coincidences (p-$\gamma$-$\gamma$) and the other employing complete kinematic measurements $via$ the $^{12}$C(p, p$\prime$)$^{12}$C reaction.
\begin{figure}
\centering
\includegraphics[scale=0.28,clip=true]{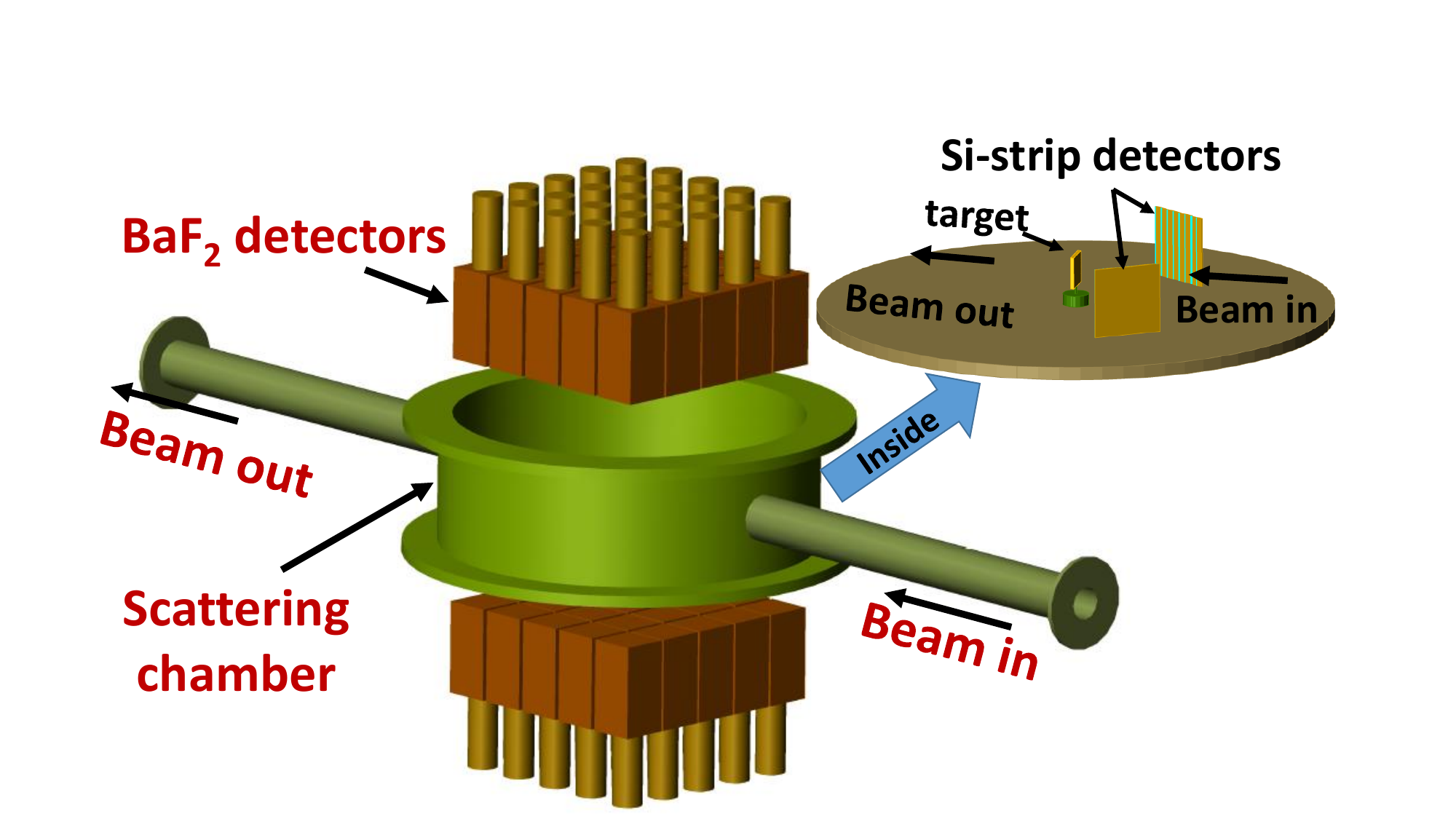}
\caption{\label{fig1} Experimental setup for the p-$\gamma$-$\gamma$ measurement.}
\end{figure}

 The experiments were performed at the Variable Energy Cyclotron Centre, Kolkata, using proton beams from the K130 cyclotron on self-supported natural carbon targets. The p-$\gamma$-$\gamma$ experiment was performed with 10.6 MeV proton beam on a carbon target (thickness $\sim$ 800 $\mu$g/cm$^2$) using $^{12}$C(p, p$\prime$$\gamma$$\gamma$)$^{12}$C reaction. The inelastic protons corresponding to the Hoyle state were detected using two silicon strip telescopes ~\cite{sknim}, while the 3.21 MeV and 4.44 MeV $\gamma$-rays, following the cascade $\gamma$-decay of the Hoyle state, were detected using a 50-element BaF$_2$ detector array~\cite{dip}. The schematic diagram of the experimental setup is shown in Fig.~\ref{fig1}. Two silicon strip telescopes were placed at backward angles (centered $\sim$125$^0$) on either side of the beam axis at a distance of $\sim$8.0 cm inside a small scattering chamber. Each strip telescope consists of a $\sim$20 $\mu$m single-sided silicon strip detector (SSSD, active area 50 mm by 50 mm, divided into 16 vertical strips, each strip of dimension 3 mm by 50 mm) as $\Delta$E and a $\sim$1030 $\mu$m double-sided silicon strip detector (DSSD, active area 50 mm by 50 mm, divided into 16 horizontal strips and 16 vertical strips; 256 pixels, each of dimension 3 mm by 3 mm) as $E$ detectors. The 50-element BaF$_2$ array (each of dimension 3.5 cm $\times$ 3.5 cm $\times$ 5.0 cm) was split into two blocks of 25 detectors. They were placed on the top and bottom of the scattering chamber at a distance of 7.5 cm from the target. The photopeak efficiency of the array was 6$\%$ at 1.78 MeV. The master trigger was taken from the OR of the two $\Delta$E detectors. Typical energy of inelastic proton at 125$^0$ corresponding to the ground, 4.44 MeV and 7.65 MeV states of $^{12}$C are 8.06 MeV, 4.23 MeV and 1.52 MeV, respectively. The trigger threshold were set ($\sim$ 0.32 MeV in $\Delta$E detectors) above the energy loss of an inelastically scattered proton from 4.44 MeV state in $^{12}$C, which assured the master trigger for the data acquisition mostly from the Hoyle state. During the experiments, the event rate was kept $\sim$2.5 kHz with a beam intensity of $\sim$3 nA. The timing signal was generated using start from the master and stop from the individual BaF$_2$ detectors. Another experiment with a natural silicon target (thickness$\sim$1 mg/cm$^2$) was performed to estimate the coincidence $\gamma$ detection efficiency of the BaF$_2$ array using the 4.98 MeV state (100 $\%$ decay $via.$ two cascade $\gamma$s of energies 1.78 and 3.20 MeVs, respectively) and the 6.28 MeV state (88.2$\%$ cascade $\gamma$ decay with $\gamma$ energies 4.50 and 1.78 MeVs) of $^{28}$Si.

 The complete kinematical experiment was performed with an 11 MeV proton beam on a carbon target (thickness $\sim$ 20 $\mu$g/cm$^2$). A Si-strip telescope [SSSD ($\sim$20$\mu$m) + DSSD ($\sim$1038 $\mu$m)], was placed at backward angle (centered at $\sim$ 120$^0$, at a distance $\sim$9 cm upstream) to detect the inelastic protons. The recoil $^{12}$C was detected using a single DSSD detector of thickness $\sim$525 $\mu$m at a forward angle (centered at 18$^0$ at a distance $\sim$25 cm downstream) in coincidence with the inelastic proton. The master trigger during the data collection was generated with a coincidence between the E and the $\Delta$E detectors of the proton telescope. In this case, the trigger threshold of the proton telescope ($\Delta$E detector) was also optimized to reduce the large contribution of scattered protons from the ground and the first excited states of $^{12}$C. A rectangular slit of 50 mm x 20 mm (made up of stainless steel with a thickness $\sim$5 mm) was placed in front of the telescope to enhance the chance of collection of the coincidence (Hoyle state) events of interest relative to other coincidence events in the raw event-by-event data. The event rate was kept at $\sim$0.2 kHz using a beam intensity of 5 nA to minimize the background noise resulting from the pile-up and the accidental coincident events. The coincident time spectrum was generated using start from the master and stop from the forward detector. For efficiency calculation of experimental setup, low threshold in $\Delta$E  detector (covering both the ground state and the 4.44 MeV state) has also been taken periodically during the experiment.  Energy calibrations of the proton telescope and forward detector were performed using a $^{229}$Th-$\alpha$ source and the elastic/inelastic peaks of the beam from $^{12}$C.

 \begin{figure}
\centering
\includegraphics[scale=0.43,clip=true]{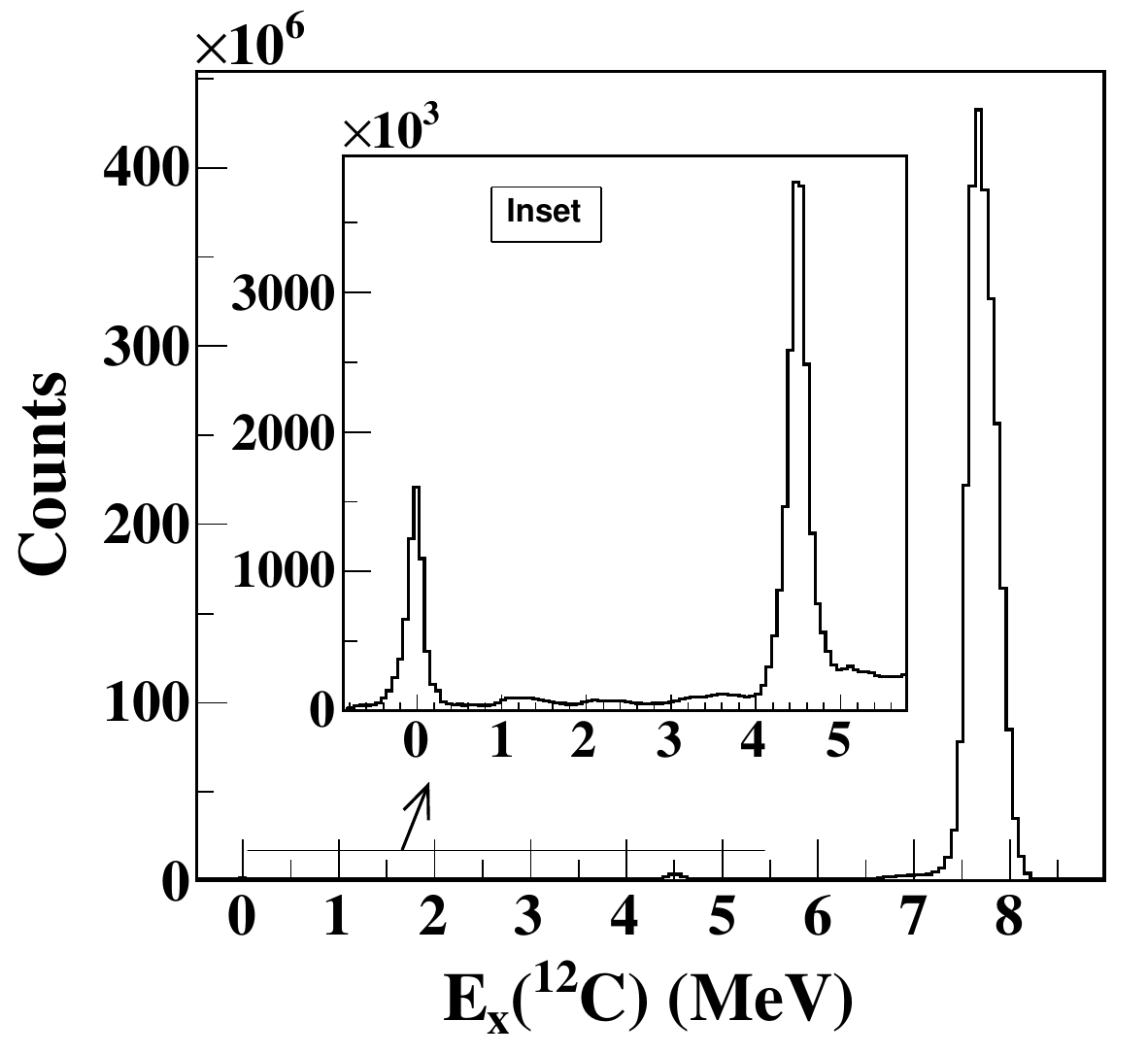}
\caption{\label{fig2} Excitation energy spectrum of $^{12}$C reconstructed from the proton energy for $^{12}$C(p, $\gamma$$\gamma$p$^{\prime}$)$^{12}$C reaction. Inset is the expanded view of the lower excitation region.}
\end{figure}
\begin{figure}
\centering
\includegraphics[scale=0.6,clip=true]{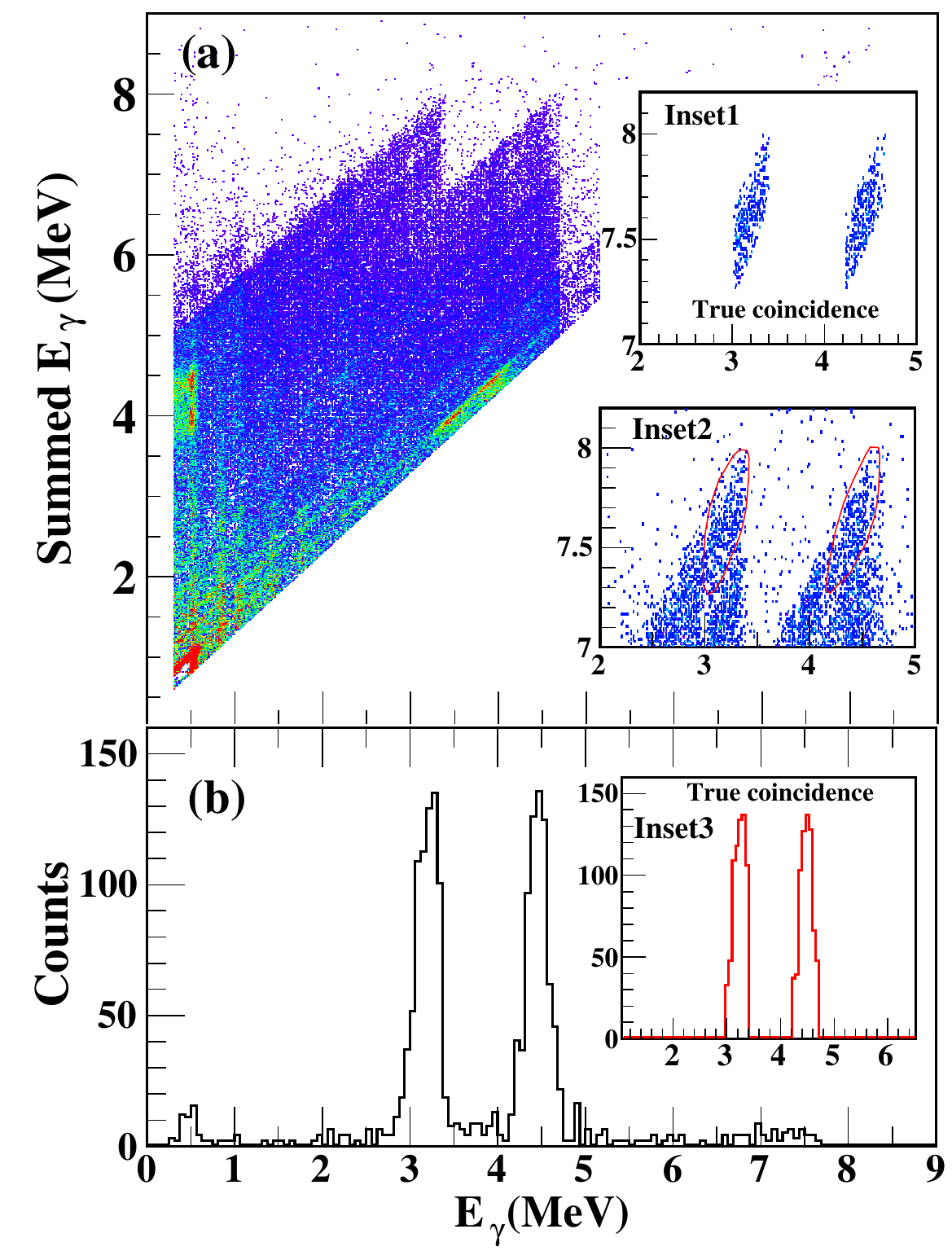}
\caption{\label{fig3} (color online) (a) Two-dimensional plot of $\gamma$-ray energy versus summed $\gamma$-ray energy gated by protons exciting the Hoyle state. Inset1 is the true coincidence 2D spectrum of 3.21 MeV and 4.44 MeV $\gamma$-rays (see text). Inset2 is the expanded view around the summed $\gamma$ energy of the Hoyle state. (b) Singles energy spectrum generated by optimizing the summed energy gates to reproduced the true coincidences (Inset3). Inset3 is the 1D projection of Inset1.}
\end{figure}
 The excitation energy spectrum of $^{12}$C reconstructed from the inelastic protons for the p-$\gamma$-$\gamma$ experiment, collected over 10 days, is shown in Fig.~\ref{fig2}. The peak at 7.65 MeV representing the Hoyle state contains 1.93 (2) $\times$ 10$^9$ events. The effect of trigger conditions is evident in the filtration of raw data, as the Hoyle state is populated 2 orders of magnitude higher than the 4.44 MeV state. A two-dimensional plot representing the $\gamma$-$\gamma$ coincident events gated by the inelastic proton exciting the Hoyle state and the prompt time gate is shown in Fig.~\ref{fig3}a. The aforementioned pair of $\gamma$ rays were measured in coincidence, focusing solely on events where one $\gamma$ was detected by the top array and the other by the bottom array. The rate of random coincidence of two 4.44 MeV $\gamma$ rays produced by two uncorrelated events and yet be detected within the prompt time gate was estimated to be 2.5 x 10$^{-5}$ /sec. This is about 30 times lower than the actual coincidence rate. The true photopeak coincidences, between 3.21 MeV and 4.44 MeV $\gamma$-rays, were determined by setting gates at $\pm$ 2.5$\sigma$ (Inset1 and Inset3 of Fig.~\ref{fig3}a). The loci of the true coincidences are indicated by the red solid lines in Inset 2. Apart from the true coincidences, the coincidences between the 4.44 MeV and the 2.7 MeV (escape peak of 3.21 MeV) and the coincidences between the 3.21 MeV and the 3.93 MeV (escape peak of 4.44 MeV) are also discernible. It is evident that the photopeak coincidences distinctly trace a different path compared to those involving coincidences between photopeak and Compton scattered or escape peaks. These three bands partially overlap with each other (on the y-axis) and cannot be fully separated using only summed energy gate. Ideally, the actual summed energy gate for selecting the true coincidence events (without the $\pm$2.5$\sigma$  gate) should range from 7.3 to 8.0 MeV, as shown in Inset 1. However, applying this summed energy gate would include coincidences with Compton scattered or escape peaks, which would increase the counts under 3.21 MeV and 4.44 MeV peaks. Therefore, the summed energy gate (without the $\pm$2.5$\sigma$  gate) was varied to reproduce the true coincidence events (obtained using only the $\pm$2.5$\sigma$ gate). An excellent match was achieved when the gate was set from 7.47 to 8.0 MeV. The singles energy spectrum, acquired by implementing only the summed energy gate, is shown in Fig. 3b. The loss of true coincidence events due to the narrower summed energy gate was estimated to be 7$\%$. This ratio was obtained by applying the summed energy gate in Inset 1 and comparing it with the spectrum without the summed energy gate (total spectrum in Inset 1), which resulted in 93$\%$. Nevertheless, since an excellent match is achieved between the summed energy spectrum and the $\pm$2.5$\sigma$  gate spectrum, the 7$\%$ loss of events are compensated by coincidences between photopeak of one $\gamma$-ray and the Compton scattered or escape peaks of other $\gamma$-ray and vice versa. The summed energy spectrum was generated to estimate the random coincidences beneath the true coincidence peaks. Finally, the counts under the peaks of 3.21 MeV and 4.44 MeV, from the summed energy spectrum (Fig.~\ref{fig3}b), were obtained as 658(27) and 649 (28), respectively by background subtracted Gaussian fitting. Similar to ref~\cite{kibedi}, we also estimated the count of inelastic protons exciting the Hoyle state as 666 (29) by gating on the prompt time, 3.21 MeV and 4.44 MeV $\gamma$s. On the other hand, by gating on the inelastic proton in the Hoyle state, along with the 3.21 MeV and 4.44 MeV $\gamma$s, the count in the prompt time was obtained as 685 (31). The average value of all the above, 665 (29) counts, was finally used for the estimation of the $\gamma$ decay width. The statistics obtained in our experiment is three times greater than those of the recent measurement ~\cite{kibedi}.

The value of the relative cascade $\gamma$ decay width of the Hoyle state can be estimated by the relation

\begin{equation}\label{3}
\frac{\Gamma_\gamma(E2)}{\Gamma}= \frac{N^{7.65}_{020}}{N^{7.65}_{singles} \times \epsilon_{3.21} \times \epsilon_{4.44} \times W^{7.65}_{020}}
\end{equation}
where $\epsilon_{3.21}$ and $\epsilon_{4.44}$ are the photopeak efficiencies (top or bottom array) for the 3.21 MeV and 4.44 MeV $\gamma$s, respectively. The symbol N$^{7.65}_{singles}$ represents the inelastic proton counts corresponding to the Hoyle state,  N$^{7.65}_{020}$  is total coincident counts of the cascade $\gamma$ decays and W$^{7.65}_{020}$ is the angular correlation correction factor for a 0-2-0 cascade $\gamma$ decay of the Hoyle state. The values of $\epsilon_{3.21}$, $\epsilon_{4.44}$ and  W$^{7.65}_{020}$ were obtained from Monte Carlo simulation based on GEANT4.

\begin{table}[ht]
\caption{Various values of experimental and Monte Carlo simulation parameters used for $\Gamma_{rad}/\Gamma$ estimation. $^*$ The efficiencies of the top and bottom arrays are identical.}
\centering 
\begin{tabular}{c  l  l l } 
\hline 
 & $^{12}$C & $^{28}$Si & $^{28}$Si \\
Parameters & 7.65 MeV & 4.98 MeV & 6.28 MeV \\
 & (020) & (020) & (320) \\
\hline 
 N$_{single}$ & 1.93(2) $\times$ 10$^9$ & 4.12(1) $\times$ 10$^5$ & 2.98(4) $\times$ 10$^6$ \\
 &  &  &  \\
  N$_{020/320}$ & 665(29) & 626(26) & 2664(53) \\
   &  &  &  \\
  $\gamma$-ray & $\epsilon_{3.21}$=2.230(7) & $\epsilon_{3.20}$=2.231(7) & $\epsilon_{4.5}$=1.723(7) \\
  efficiency($\%$) & $\epsilon_{4.44}$=1.747(7) & $\epsilon_{1.78}$=3.054(7) & $\epsilon_{1.78}$=3.054(7) \\
   (top/bottom)$^*$ &  &  &  \\
      &  &  &  \\
  W$_{020/320}$ & 1.1180(30) & 1.1089(30) & 0.9834(30)\\
\hline 
\end{tabular}
\label{table1}
\end{table}

To check the accuracy of the simulation, the 1.78 MeV, 4.98 MeV, and 6.28 MeV states of $^{28}$Si were utilized. The simulation was fine-tuned using the single energy $\gamma$-ray from the proton-gated spectrum obtained from the 1.78 MeV state. Subsequently, the coincidence measurements were validated with the spectra from the 4.98 MeV and 6.28 MeV states. The same number of events were generated as measured in the experiments (considering the angular correlation). The coincidence counts obtained under the peak of 1.78 MeV and 3.20 MeV $\gamma$-rays from the 4.98 MeV state, by Gaussian fitting, were 624 and 632, respectively, while for 1.78 MeV and 4.50 MeV (6.28 MeV state) they were obtained as 2972 and 3013, respectively. Considering, the 88.2 $\%$ branching ratio for the 6.28 MeV state, the average count under the peaks of 1.78 MeV and 4.50 MeV was obtained as 2639, in excellent agreement with the experimental data (Table~\ref{table1}). Utilizing the values from Table~\ref{table1} and taking into account the BaF$_2$ array combinations as 2, $\Gamma_\gamma(E2)/\Gamma$ was found to be 3.95 (28) $\times$ 10 $^{-4}$.

The $\Gamma_\gamma$(E2)/$\Gamma$ has also been estimated using the 4.98 MeV state of $^{28}$Si by the relation as done in previous works ~\cite{kibedi}
\begin{equation}\label{4}
\frac{\Gamma_\gamma(E2)}{\Gamma}= \frac{N^{7.65}_{020}}{N^{4.98}_{020}}\times \frac{N^{4.98}_{singles}}{N^{7.65}_{singles}} \times \frac{\epsilon_{1.78}}{\epsilon_{4.44}}\times \frac{\epsilon_{3.20}}{\epsilon_{3.21}} \times \frac{ W^{4.98}_{020}}{ W^{7.65}_{020}}
\end{equation}

This relation reduces the dependency on the simulation as well as the angular correlation. The $\Gamma_\gamma(E2)/\Gamma$ was obtained from Eq.\ref{4} by using the values from Table~\ref{table1} as 3.92 (28) $\times$ 10 $^{-4}$, which is consistent with the value obtained using Eq.~\ref{3}. Using the theoretical total conversion coefficient of a 3.21 MeV E2 transition, 8.77 (13) $\times$ 10$^{-4}$ ~\cite{kibedi1}, and the recommended value of $\Gamma_\pi$(E0)/$\Gamma$~\cite{kelley}, we obtain $\Gamma_{rad}/\Gamma$ = 4.01 (30) $\times$  10$^{-4}$. The value of $\Gamma_{rad}/\Gamma$ obtained recently~\cite{kibedi} is  55 $\%$ higher than our experimental value.
\begin{figure}
\centering
\includegraphics[scale=0.43,clip=true]{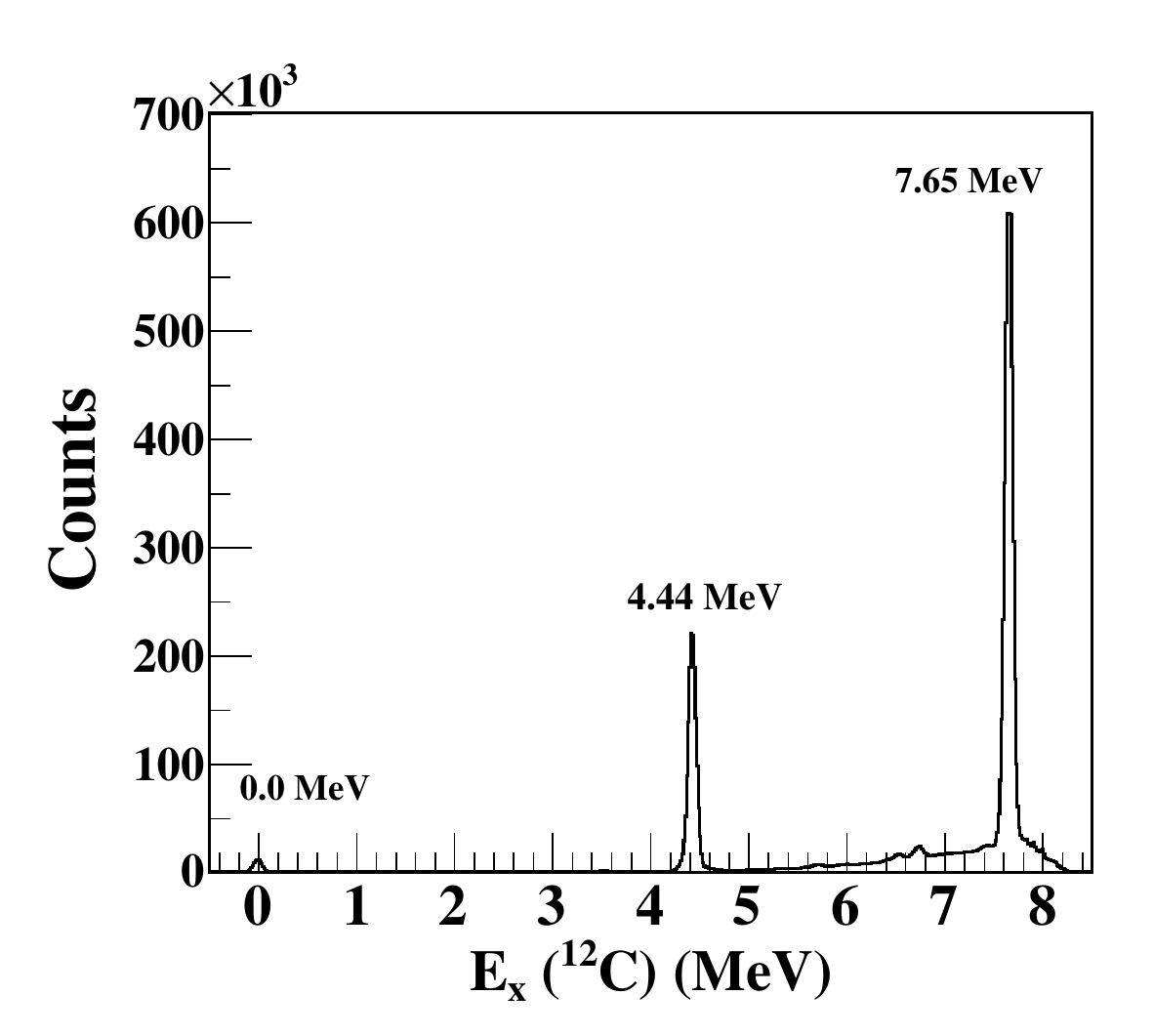}
\caption{\label{fig4} (color online) Excitation energy of $^{12}$C reconstructed from the inelastic proton energy for the complete kinematical measurement in $^{12}$C(p, p$^{\prime}$)$^{12}$C reaction.}
\end{figure}
\begin{figure}
\centering
\includegraphics[scale=0.43,clip=true]{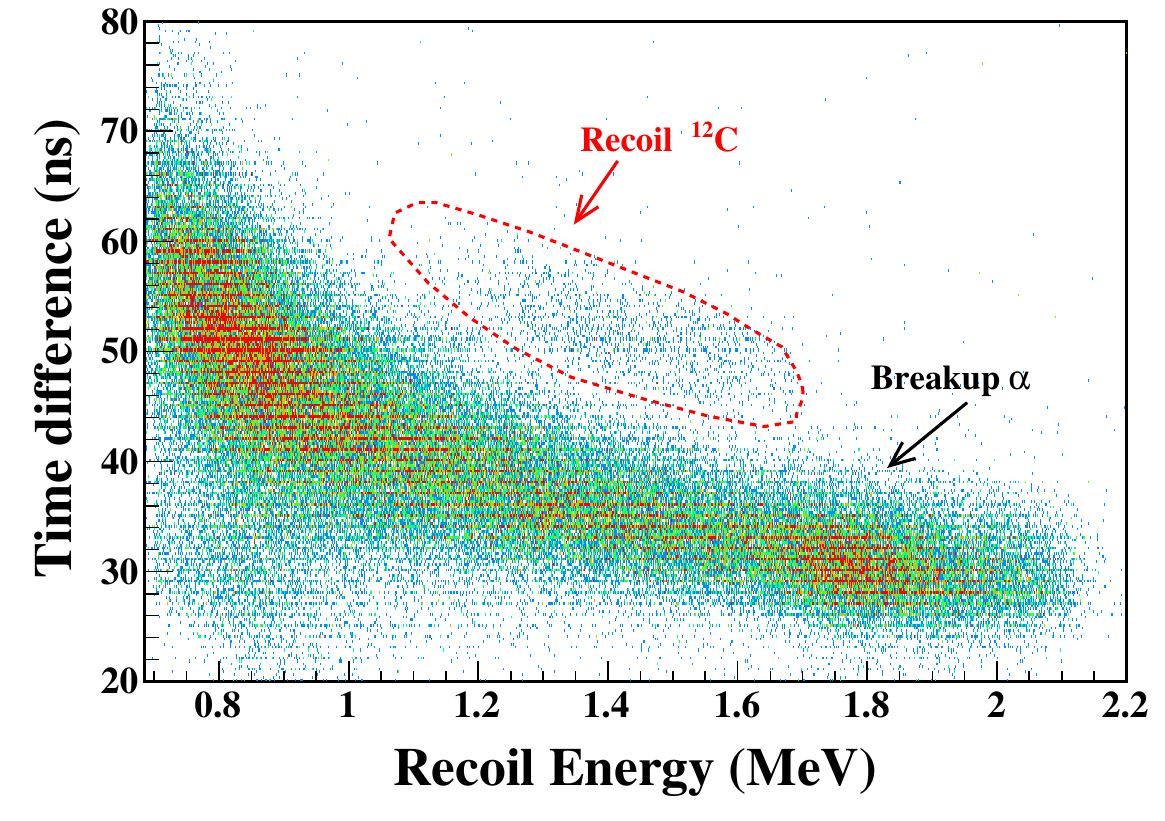}
\caption{\label{fig5} (color online) Plot between the relative time of flight and the forward detector energy with a gate on inelastic proton corresponds to the Hoyle state.}
\end{figure}

The $\Gamma_{rad}$/$\Gamma$ has also been measured from a complete kinematical experiment using the reaction $^{12}$C(p, p$^{\prime}$)$^{12}$C. The excitation energy of $^{12}$C reconstructed from the inelastic proton is shown in Fig.~\ref{fig4}. As can be seen, the population of the Hoyle state is dominant compared to the 4.44 MeV or ground state of $^{12}$C because of the threshold optimization of the $\Delta$E-detector of the proton telescope. The total number of Hoyle states excited by the proton was found to be 4.37 (2) × 10$^6$.

Since the Hoyle state decays back to the same entrance channel~\cite{rana1,rana}, the break up $\alpha$s are also detected along with the
recoil $^{12}$C (formed after radiative decay of the Hoyle state) in the forward detector. To distinguish them, the relative time of flight of the detected particles, with respect to the inelastic proton, was plotted against the energy recorded in the forward detector (see Fig.~\ref{fig5}). The time resolution for single energy carbon and inelastic proton pair was $\sim$ 4 ns. A clear separation is obtained between the recoil $^{12}$C and the breakup $\alpha$s corresponding to the Hoyle state (Fig.~\ref{fig5}). It is to be noted that the energy correction for the detector dead layer was performed for all particles detected in the forward detector under the assumption they were all $^{12}$C. Hence, the energy of breakup $\alpha$s (supposed to be maximum energy of 1.2 MeV within the angular coverage of the forward detector) is higher compared to the recoil $^{12}$C energy. The random background in the two-dimensional spectrum (Fig.~\ref{fig5}) was obtained by putting a gate just below the Hoyle state in the inelastic proton and then subtracted. However, the decrease in recoil $^{12}$C counts was found to be less than 1$\%$. The total number of recoil $^{12}$C, formed after radiative decay, was obtained as 1501 (39), collected over 5 days. The ratio of the experimental counts of recoil $^{12}$C to the total inelastic excitations with efficiency correction results $\Gamma_{rad}$/$\Gamma$.

The complete kinematical detection efficiencies of the setup for detecting the recoil $^{12}$C from the ground and 4.44 MeV states of $^{12}$C were measured periodically during the experiment. As it cannot be measured directly for the 7.65 MeV state, the efficiency of the experimental setup was obtained from Monte Carlo simulation by considering the experimental effects ~\cite{rana1,rana}. A detailed Monte Carlo simulation was performed to reproduce the experimental efficiencies corresponding to the ground and 4.44 MeV states. The agreement between the calculated and the measured efficiencies was within 98.5 $\%$. Next, the efficiency of the setup for the Hoyle state was obtained using the same Monte Carlo simulation. The calculated efficiencies for detection of recoil $^{12}$C in the ground state, 4.44 MeV, and 7.65 MeV states were 15 $\%$, 45$\%$, and 85$\%$, respectively. The value of $\Gamma_{rad}$/$\Gamma$, after efficiency correction was found to be 4.04 (30) $\times$ 10$^{-4}$, which is consistent with the previous value measured in  $^{12}$C(p, p$^\prime\gamma$$\gamma$)$^{12}$C reaction. By averaging the two experimental values of $\Gamma_{rad}$/$\Gamma$ (= 4.03 (21) $\times$ 10$^{-4}$) combined with the recommended pair decay and relative pair decay widths ~\cite{kelley}, $\Gamma_{rad}$, was found to be  3.75 (40) $\times$ 10$^{-3}$ eV. The uncertainty in the estimated value includes both the statistical and the systematic errors. As could be discerned, the present study does not reproduce the 34 $\%$ increase in the radiative decay width of the Hoyle state proposed recently~\cite{kibedi}. Instead, it aligns with all previous measurements \cite{freer,kelley}, as well as with the recent measurements during the examination of the 3$^-$ state $\gamma$ decay of carbon ~\cite{tsumura} and charged particle spectroscopy ~\cite{luo}. The present value of $\Gamma_{rad}$/$\Gamma$, when combined with the latest determination of $\Gamma_\pi(E0)$/$\Gamma$ by T. K. Eriksen et al.~\cite{eriksen} (showing a 14$\%$ increase compared to the prior value), yields a new value for $\Gamma_{rad}$ of 3.30(26) × 10$^{-3}$ eV. Interestingly, this value is even lower than the one reported recently ~\cite{kibedi}.

In conclusion, two independent measurements were performed to estimate the radiative decay branching ratio of the Hoyle state, using the reaction $^{12}$C(p, p$\prime$)$^{12}$C. A triple coincidence measurement of p-$\gamma$-$\gamma$ in the exit channel yielded a value of $\Gamma_{rad}$/$\Gamma$ = 4.01 (30) $\times$ 10$^{-4}$. This result was further confirmed by a complete kinematic measurement, which gave a value of $\Gamma_{rad}$/$\Gamma$ = 4.04 (30) $\times$ 10$^{-4}$. The average value of $\Gamma_{rad}$/$\Gamma$ was obtained as 4.03 (21) $\times$ 10$^{-4}$ and the corresponding radiative width of the Hoyle state was determined to be 3.75 (40) $\times$ 10$^{-3}$ eV. The radiative decay width determined in current work is notably smaller than the recent measurement ~\cite{kibedi}. The present result is both significant and novel, as it, for the first time, incorporates two types of analysis—$\gamma$-ray spectroscopy and charged particle spectroscopy—based on two independent measurements. Our result reaffirms the previously adopted value~\cite{freer,kelley} and can be seamlessly applied to astrophysical reaction rate calculations.

The authors are thankful to the K130 cyclotron crew for the smooth operation of the machine during the experiments. The authors like to thank J. K. Meena, J. K. Sahoo, A. K. Saha and R. M. Saha for their help during the experimental setup and data taking. The authors are also thankful to Shri Partha Dhara and his team for their support in data acquisition system.


\begin{thebibliography}{50}
\bibitem{hoyle} F. Hoyle, Astrophys. J. Suppl. Ser. \textbf{1}, 12 (1953).
\bibitem{dubnar} D. N. F. Dunbar, R. E. Pixley, W. A. Wenzel, and W. Whaling, Phys. Rev. \textbf{92}, 649 (1953).
\bibitem{BFH} E. Margaret Burbidge, G. R. Burbidge, Willian A. Fowler and F. Hoyle, Rev. of Mod. Phys. \textbf{29}, 4 (1957).
\bibitem{fowler} W. A. Fowler Rev. Mod. Phys. 56, 149–179 (1984).
\bibitem{beard} M. Beard, S. M. Austin,  R. Cyburt, Phys. Rev. Lett. \textbf{119}, 112701 (2017).
\bibitem{shilun} Shilun Jin, Luke F. Roberts, Sam M. Austin and Hendrik Schatz, Nature \textbf{588}, 57 (2020). 
\bibitem{wanajo} S. Wanajo, H.-T. Janka, S. Kubono, Astrophys. J. \textbf{729}, 46 (2011).
\bibitem{woosley} S. E. Woosley and A. Heger, Astrophys. J. Lett. \textbf{912}, L31 (2021).
\bibitem{rana1} T. K. Rana, S. Bhattacharya,  C. Bhattacharya, S. Manna, S. Kundu, K. Banerjee $et$ $al.$ Phys. Lett. B \textbf{793}, 130 (2019).
\bibitem{Dell} D. DellAquila, I. Lombardo, G. Verde, M. Vigilante, L. Acosta, C. Agodi $et$ $al.$, Phys. Rev. Lett. \textbf{119}, 132501 (2017).
\bibitem{Smith} R. Smith, Tz. Kokalova, C. Wheldon, J. E. Bishop, M. Freer, N. Curtis and D. J. Parker, Phys. Rev. Lett. \textbf{119}, 132502 (2017).
\bibitem{freer} M. Freer and H. O. U. Fynbo, Prog. Part. Nucl. Phys. \textbf{78}, 1 (2014) and reference therein.
\bibitem{mfd} J. Manfredi, R. J. Charity, K. Mercurio, R. Shane, L. G. Sobotka, A. H. Wuosmaa, A. Banu, L. Trache, and R. E. Tribble, Phys. Rev. C \textbf{85} 037603 (2012).
\bibitem{nupscat} J. Bishop $et$ $al.$,  Nat Commun \textbf{13}, 2151 (2022).
\bibitem{rolf} C. E. Rolfs and W. S. Rodney, Cauldrons in the Cosmos: Nuclear Astrophysics (University of Chicago Press, Chicago, (1988).
\bibitem{freg}J. H. Fregeau, Phys. Rev. \textbf{104}, 225 (1956).
\bibitem{cran} H. L. Crannell and T. A. Griffy, Phys. Rev. \textbf{136}, B1580 (1964).
\bibitem{gud} F. Gudden and P. Strehl, Z. Phys. \textbf{185}, 111 (1965).
\bibitem{cran1} H. Crannell, T. A. Griffy, L. R. Suelzle, and M. R. Yearian, Nucl. Phys. A \textbf{90}, 152 (1967).
\bibitem{streh} P. Strehl and Th. H. Schucan, Phys. Lett. B \textbf{27}, 641 (1968).
\bibitem{streh1} P. Strehl, Z. Phys. \textbf{234}, 416 (1970).
\bibitem{cran2} H. Crannell, X. Jiang, J. T. \`{O}Brien, D. I. Sober, and E. Offermann, Nucl. Phys. A \textbf{758}, 399 (2005).
\bibitem{chernykh} M. Chernykh, H. Feldmeier, T. Neff, P. von Neumann-Cosel, and A. Richter, Phys. Rev. Lett. \textbf{105}, 022501 (2010).
\bibitem{selove} F. Ajzenberg-Selove and P. H. Stelson, Phys. Rev. \textbf{120}, 500 (1960).
\bibitem{alburger} D. E. Alburger, Phys. Rev. \textbf{118}, 235 (1960).
\bibitem{obst} A.W. Obst and W. J. Braithwaite, Phys. Rev. C \textbf{13}, 2033 (1976).
\bibitem{alburger1} D. E. Alburger, Phys. Rev. C \textbf{16}, 2394 (1977).
\bibitem{robertson} R. G. H. Robertson, R. A. Warner, and S. M. Austin, Phys. Rev. C \textbf{15}, 1072 (1977).
\bibitem{eriksen} T. K. Eriksen, T. Kib´edi, M.W. Reed, A. E. Stuchbery, K. J. Cook, A. Akber et al., Phys. Rev. C \textbf{102}, 024320 (2020).
\bibitem{alburger2} D. E. Alburger, Phys. Rev. \textbf{124}, 193 (1961).
\bibitem{seeger} P. A. Seeger and R.W. Kavanagh, Nucl. Phys. \textbf{46}, 577 (1963).
\bibitem{obst1} A. W. Obst, et al., Phys. Rev. C \textbf{5} (1972) 738.
\bibitem{hall} I. Hall, N.W. Tanner, Nuclear Phys. \textbf{53} (1964) 673.
\bibitem{chamber} D. Chamberlin, et al., Phys. Rev. C \textbf{9} (1974) 69.
\bibitem{david} C. N. Davids, et al., Phys. Rev. C \textbf{11} (1975) 2063.
\bibitem{mak} H. B. Mak, et al., Phys. Rev. C \textbf{12} (1975) 1158.
\bibitem{markh} R. G. Markham, S.M. Austin, M.A.M. Shahabuddin, Nuclear Phys. A \textbf{270} (1976) 489.
\bibitem{kelley} J. Kelley, J. Purcell, and C. Sheu, Nucl. Phys. A \textbf{968}, 71 (2017).
\bibitem{kibedi} T. Kib\'{e}di, B. Alshahrani, A. E. Stuchbery, A. C. Larsen, A. G\"{o}rgen, S. Siem et. al, Phys. Rev. Lett. \textbf{125}, 182701 (2020).
 \bibitem{sknim} Samir Kundu $et$ $al.$, Nucl. Inst. Meth. A \textbf{943}, 162411 (2019).
\bibitem{dip} Deepak Pandit, S. Mukhopadhyay, Srijit Bhattacharya, Surajit Pal, A. De, S. R. Banerjee, Nucl. Instrum. Methods A \textbf{624}, 148 (2010).
\bibitem{kibedi1} T. Kib\'{e}di, T.W. Burrows, M. B. Trzhaskovskaya, P. M. Davidson, and C. J. Nestor, Nucl. Instrum. Methods Phys. Res., Sect. A \textbf{589}, 202 (2008).
 \bibitem{rana} T. K. Rana, S. Bhattacharya, C. Bhattacharya, S. Kundu, K. Banerjee, T. K. Ghosh, G. Mukherjee $et$ $al.$ Phys. Rev. C \textbf{88}, 021601(R) (2013).
 \bibitem{tsumura} M. Tsumura $et$ $al.$ Phys. Lett. B \textbf{817}, 136283 (2021).
\bibitem{luo} Zifeng Luo $et$ $al.$, Phys. Rev. C \textbf{109}, 025801 (2024).
\end{thebibliography}
\end{document}